\documentclass[aps,showpacs,preprint,groupedaddress]{revtex4}
\usepackage{graphics}
\usepackage{graphicx}
\usepackage{psfig}
\usepackage{epsf}
\usepackage{amsmath}
\usepackage{amssymb}
\usepackage{bm}
\usepackage{dcolumn}
\begin{document}
\title{Phase and Charge reentrant phase transitions in two capacitively coupled
Josephson arrays with ultra-small junction}
\author{Guillermo Ram\'{\i}rez-Santiago}
\affiliation{Instituto de F\'{\i}sica, Universidad Nacional\\
Aut\'onoma de M\'exico, Apartado Postal 20-364,\\
M\'exico 01000, D. F., MEXICO}
\author{Jorge V. Jos\'e}
\affiliation{Department of Physics and Center for Interdisciplinary\\
Research on Complex Systems, Northeastern University, 360\\
Huntington Ave. Boston MA 02115, USA}
\begin{abstract}
We have studied the phase diagram of two capacitively coupled
Josephson junction arrays with charging energy, $E_c$, and
Josephson coupling energy, $E_J$. Our results are obtained using a
path integral Quantum Monte Carlo algorithm. The parameter
that quantifies the quantum fluctuations in the i-th array is defined by $%
\alpha_i\equiv \frac{E_{{c}_i}}{E_{J_i}}$. Depending on the value of $%
\alpha_i$, each independent array may be in the semiclassical or
in the quantum regime: We find that thermal fluctuations are
important when $\alpha \lesssim 1.5 $
and the quantum fluctuations dominate when $2.0 \lesssim \alpha $.
Vortices are the dominant excitations in the semiclassical limit,
while in the quantum regime the charge excitations are important.
 We have extensively studied the interplay between vortex
and charge dominated individual array phases. The phase diagrams
for each array as a function of temperature and inter-layer
capacitance where determined from results for their helicity
modulus, $\Upsilon (\alpha)$, and the inverse dielectric constant,
$\epsilon^{-1} (\alpha)$. The two arrays are coupled via the
capacitance $C_{{\rm inter}}$ at each site of the lattices. When
one of the arrays is in the quantum regime and the other one is in
the semi-classical limit, $\Upsilon(T,\alpha)$ decreases with $T$,
while $\varepsilon^{-1}(T,\alpha) $ increases. This behavior is
due to a duality relation between the two arrays:  e. g. a
manifestation of the {\it gauge invariant capacitive interaction}
between vortices in the semiclassical array and charges in the
quantum array. We find a {\it reentrant transition} in
$\Upsilon(T,\alpha)$,  at low temperatures, when one of the arrays
is in the semiclassical limit (i.e. $\alpha_{1}=0.5 $) and  the
quantum array has $2.0 \leq\alpha_{2} \leq 2.5$, for the values
considered for  the interlayer capacitance of $C_{{\rm inter}}=
0.26087, 0.52174, 0.78261, 1.04348$ and $1.30435$. Similar results
were obtained for larger values of $\alpha_{2} =4.0$ with $C_{{\rm
inter}}=1.04348$ and 1.30435. For smaller values of $C_{{\rm
inter}}$ the superconducting-normal transition was not present. In
addition, when $3.0 \leq \alpha_{2} < 4.0$, and for all the
inter-layer couplings considered above, a {\it novel} reentrant
phase transition occurs in the charge degrees of freedom, i.e.
there is a reentrant insulating-conducting transition at low
temperatures. Finally, we obtain the corresponding phase diagrams
that have some features that resemble those seen in experiment.
\end{abstract}


\maketitle
\newpage
\section{introduction}

\label{introduction}

This paper considers the phase diagrams of two capacitively
coupled Josephson junction arrays (JJA), made of ultrasmall
junctions. Two-dimensional JJA have been the subject of many
theoretical~\cite
{jose84,jacobs84,jacobs87,jacobs88,kim90,fazio91,simkin91,ariosa92,granato92,granato93,
roddick93,jose94,simakek94,kim95,rojas96,rojas-thesis,kopec2000,luca2003,luca2004}
and experimental~\cite
{fazio_herre,sohn,Takahide,Haviland2000,proceed,mooij90,herre91,herre92,tighe93,delsing94,elion95,herre96}
studies. Advances in submicrometer
technology~\cite{fazio_herre,sohn} and  in nanolitographic
techniques~\cite{Takahide,Haviland2000} have allowed the
fabrication of relatively large arrays made of ultrasmall
superconductor-insulator-superconductor Josephson junctions (JJ).
The JJ areas may vary from a few microns to submicron dimensions,
with self-capacitances $C_{s} \approx 3\times 10^{-2}$ fF and
nearest neighbor mutual capacitance $C_{m} \approx 1$ fF. Notice
that the mutual capacitance can be at least two orders of
magnitude larger than the self-capacitance.

In these arrays  there are two competing energies: the Josephson
coupling energy, $E_{J}$, and the charging energy
$E_{C}=\frac{e^{2}}{2C_{m}}$ of the junctions, due to the charge
localization in the islands, where $e$ is the electronic charge.
In the semiclassical limit, $E_{J}>>E_{C}$, the phase of the
superconducting order parameter of the junctions is well defined.
The Josephson coupling induces fluctuations in the charge number
that produces a supercurrent, where the average Cooper pair number
is undefined. In this regime the vortex excitations are pinned by
the intrinsic lattice potential and the array is in a
superconducting state. In the quantum limit, $E_{J}<<E_{C}$, the
electrostatic energy to add one Cooper pair in one of the two
islands in a junction is much larger than the thermal energy. The
electric field localizes the Cooper pairs in the islands and the
quantum fluctuations of the phase of the superconducting order
parameter delocalizes the vortex excitations. This charge
localization in the junction islands drives the array to an
insulating state. The number of Cooper pairs, $n$, and the phase
of the superconducting order parameter, $\phi $, obey the
Heisenberg's uncertainty principle, $\Delta n\Delta \phi
\geq{\dfrac{1}{2}}$.
This uncertainty relation has been demonstrated experimentally in ${\rm Al-Al}_{2}{\rm O}%
_{3}-{\rm Al}$ junctions ~\cite {mooij-nature}.

The superconducting-insulating (S-I) phase transition, induced by
the charging energy in arrays of this type, has been
experimentally measured as a function of $\alpha$ by groups in
Delft~\cite{herre92} and Harvard~\cite{tighe93}. Their junction
sizes had constant values, while they varied the normal state
junction resistance to change the Josephson coupling energy. This
allowed them to fabricate arrays with $\alpha $ in the range
$[0.13-4.55]$~\cite{herre92}, or as high as 33~\cite{tighe93}. JJA
have also been studied in connection with quantum phase
transitions~\cite{kopec2000}. The JJA have the advantage, over
films, that their internal structure can be carefully controlled
experimentally. The drawback is that the array sizes are limited
by fabrication constraints. There is a large literature studying
the phase transition structure of one JJA with ultra small
junctions as the temperature, the external magnetic field and
$\alpha$ are varied (for a comprehensive recent review see
\cite{fazio_herre}).

A novel  experimental system composed of {\it two capacitively
coupled JJA} made of ultrasmall junctions was discussed in a paper
by the Delft group \cite{sohn}, in which each array was produced
with different $\alpha$ parameters. Initial theoretical studies of
this system were started in \cite{rojas95}, \cite{blatter} and in
\cite{jose98}. In this paper we study the phase diagram of this
very interesting system. The Hamiltonian describing the coupled
arrays can be formally written as  $\hat{\cal H}={\hat
H}_J(1)+{\hat H}_J(2)+ {\hat H}_C({1,2})$, where the ${\hat
H}_J(i)$ denote the Josephson Hamiltonians for each array, and
$H_C(1,2)$ describes the two arrays capacitive interactions.
$H_C(1,2)$ includes the total charging energy matrix, which
includes the self- and mutual capacitive terms in each plane, plus
the arrays interaction due to the ultrasmall nearest neighbor
capacitive coupling between them. This Hamiltonian is more
complicated to study than the single array problem studied
before~\cite{herre92,fazio_herre}. Here each array is described by
the ratio $\alpha_i =(E_{C_i}/E_{J_i})$ ($i=1,2$), with the mutual
array capacitive coupling denoted by $\alpha_{int}$. In the
$\alpha_i\ll 1$ limit, the $i-th$ array is dominated by localized
vortex excitations, $V_i$, while the Cooper pair excess charge
excitations, $Q_i$, are in a superconducting state. In the $\alpha
_i\gg 1$ regime, the system has the $Q_i$'s localized in an
insulating state while the $V_i$'s are delocalized. There are
different parameter regimes that can be studied. Here we consider
in detail the case when one array is $V$ dominated and the other
array $Q$ dominated. Using the Villain approximation and duality
transformations~\cite{jose77}, it was
 shown~\cite{rojas95,blatter,jose98} that the interaction
between vortices and charges has a minimal gauge coupling form,
and that the interaction is sharply localized in space:  A vortex and a
Cooper pair interact only  if they are located at the same site in
both arrays. In previous studies an effective Hamiltonian was
derived, in terms of $V_1$ and $Q_2$, which shows a high degree of
dual symmetry, with the $V_1 \iff Q_2$ interaction expressed as
{\it a capacitively induced minimal gauge like coupling}
\cite{rojas95,blatter,rojas96,jose98}.

We used a path integral Quantum Monte Carlo algorithm to further
study the behavior of this system. We calculated the phase diagram
described in terms of the vortex helicity modulus, $\Upsilon$, and
the charge inverse dielectric constant, $\epsilon ^{-1}$, as a
function of $\alpha$ and temperature. We find that as $\alpha$
varies there are different {\it novel} reentrant phase transitions
seen in $\Upsilon$ and $\epsilon ^{-1}$.  These re-entrances
indicate transitions between SC-N, SC-I and I-N phases. The layout
of the paper is as follows: In section~\ref{model} we define and
explain the model of interest and the path integral formalism to
evaluate its partition function, as well as the physical
quantities that are calculated to characterize the behavior of the
model. In section~\ref{results-discussion} we present and discuss
the results of our calculations and finally, in
section~\ref{conclusions}, we summarize our main findings and
present our conclusions.

\section{The model and its path integral representation}
\label{model}

In this section we define the model that describes
two JJA made of ultra small junctions capacitively coupled at each
lattice site. The parameter values used in the simulations
correspond to those present in the experimental samples fabricated
and studied at Delft~\cite{herre92}. The dimensions of the
experimental arrays were $L_{x}=230$ and $L_{y}=60$, with
intra-layer mutual capacitance  of $C_{{\rm m}}=2.3$ fF, and
inter-layer capacitance  $C_{{\rm int}}=0.6$ fF.

The model Hamiltonian is defined by~\cite{jose98},
\begin{eqnarray}
\label{hamiltonian}
{\cal H}&=& \frac{Q^{2}}{2}
\sum_{<\vec{r}_{1},\vec{r}_{2}>}\sum_{\mu =1,\nu =1}^{2}{\bf
n}_{\mu }(\vec{r}_{1})\>\tilde{{\bf C}}_{\mu ,\nu }^{-1}(\vec{r}_{1},\vec{r}_{2})\>{\bf n}
_{\nu }(\vec{r}_{2}) +E_{J_1}\sum_{<\vec{r}_{1},\vec{r}_{2}>}\Big(1-
\cos \big(\phi _{1}(\vec{r}_{1})-\phi _{1}(\vec{r}_{2})\big)\Big) \cr
& +&E_{J_2}\sum_{<\vec{r}_{1},\vec{r}_{2}>}\Big(1-\cos \big(\phi _{2}(\vec{r}_{1})-
\phi _{2}(\vec{r}_{2})\big)\Big),
\end{eqnarray}
where the sums are over nearest neighbors, and $\mu $ and $\nu $
label the matrix elements of the mutual capacitive interactions. $\vec{r_{1}}$ and $%
\vec{r_{2}}$ denote the positions of the junctions in arrays 1 and 2, respectively. 
The operator  $\hat{{\bf n} _{\mu}}(\vec{r}_{1})$, for the number of Cooper pairs, 
and the phase of
the superconducting order parameter $\hat{\phi _{\nu }}(\vec{r}_{2})$, satisfy the
Heisenberg commutation relations, $[\hat{%
{\bf n}_{\mu }}(\vec{r}_{1}),\hat{\phi _{\nu }}(\vec{r}_{2})]=-i\delta _{%
\vec{r}_{1},\vec{r}_{2}}\delta _{\mu ,\nu }$. The charge carried
by the Cooper pairs is $Q=2e$, and $E_{J_1}$ and $E_{J_2}$ are
the Josephson coupling constants within each array. $\tilde{{\bf C}^{-1}}%
_{\mu ,\nu }$ is the electric field propagator and its inverse, ${\bf C}%
_{\mu ,\nu }$, is the block matrix  representing the geometric
capacitance. In what follows we will use the notation ${\bf C}_{\mu ,\mu }={\bf C}%
_{\mu }$ to denote the diagonal capacitance matrix. The block
matrices are given by,

\begin{equation}
\label{block-matrix}
     {\bf C}_{\mu,\nu}(\vec r_1,\vec r_2) =
                \left\{ \begin{array}{ll}
                        \left(C_{\rm s,\mu}+zC_{\rm m , \mu}+C_{\rm int}
                        \right),
                                    & \mbox{if $\mu=\nu$ and $\vec r_1=
                                                              \vec r_2$},\\
                        -C_{\rm m ,\mu},
                                    & \mbox{if $\mu=\nu$ and $\vec r_1=
                                                        \vec r_2\pm\vec u$},\\
                        -C_{\rm int},
                                    & \mbox{if $\mu\ne\nu$ and $\vec r_1=
                                                                \vec r_2$},\\
                          0,
                                    & \mbox{otherwise.}
                        \end{array}
                \right.
\end{equation}
Here $\vec{d}$ is a unit vector and $z$ is the lattice
coordination number. The off-diagonal blocks are written as
$-C_{{\rm int}}\,{\bf I}_{\,N,N}$ with ${\bf I}$ the identity
matrix and $N$ the size of the arrays. The block matrix elements
in the Hamiltonian in Eq. ~(\ref{hamiltonian}) are written as~
\cite{jose98},

\begin{eqnarray}
\label{c-tilde-11}
    & &{\bf\tilde C}_{1,1} = {\bf C}_1^{-1}\left[{\bf I} - C_{\rm int}^2
       {\bf C}_2^{-1}{\bf C}_1^{-1}\right]^{-1},\nonumber \\
    & &{\bf\tilde C}_{2,2} = {\bf C}_2^{-1}\left[{\bf I} - C_{\rm int}^2
       {\bf C}_1^{-1}{\bf C}_2^{-1}\right]^{-1},\nonumber \\
    & &{\bf\tilde C}_{1,2} = C_{\rm int} {\bf C}_1^{-1} {\bf C}_2^{-1}
       \left[{\bf I}-C_{\rm int}^2{\bf C}_1^{-1}{\bf C}_2^{-1}\right]^{-1},\nonumber \\
    & &{\bf\tilde C}_{2,1} = C_{\rm int} {\bf C}_2^{-1} {\bf C}_1^{-1}
       \left[{\bf I}-C_{\rm int}^2{\bf C}_2^{-1}{\bf C}_1^{-1}\right]^{-1}.
\end{eqnarray}
To evaluate the partition function, $\Xi $, we can calculate the
trace over the phase operators $\hat{\phi}$ or over the trace  of the number
operators $\hat{{\bf n}}$. The path integral
representation of $\Xi $ can be obtained using the
states~\cite{kleinert},
\begin{equation}
\langle n(\vec{r_{1}})|\phi (\vec{r_{2}})\rangle =
{\displaystyle{ \frac{\exp [in(\vec{r_{1}})\phi (\vec{r_{2}})]} {\sqrt{2\pi} }}}%
\delta _{\vec{r_{1}},\vec{r_{2}}}.
\end{equation}
Following the steps outlined in reference~\cite{rojas96},  we obtain the lattice path
integral representation of the partition function,
\begin{eqnarray}
{\Large \Xi} & = &
\prod_{\tau=0}^{L_{\tau} -1 }  \prod_{\vec{r}}
\sum_{n(\tau,\vec{r})}
\int_{0}^{2\pi} \dfrac{d\phi(\tau,\vec{r})} {2\pi} \exp
\Bigg\lbrace
-\int_{0}^{\beta\hslash} d\tau
\Bigg[
\sum_{\vec{r_{1}},\vec{r_{2}}}
\dfrac{Q^{2}} {2} n(\tau,\vec{r_{1}})
{\bf C}^{-1} (\vec{r_{1}},\vec{r_{2}})
n(\tau,\vec{r_{2}}) \cr
&+& i\sum_{\vec{r}}  n(\tau,\vec{r})  \dfrac{d \phi}{d \tau} (\tau,\vec{r})
+ E_{J}\sum_{<\vec{r_{1}},\vec{r_{2}}>}
\Big[ 1 - \cos \big( \phi(\tau,\vec{r_{1}}) - \phi(\tau,\vec{r_{2}}) \big)
\Big] \Bigg]
\Bigg\rbrace.
\label{partition}
\end{eqnarray}

Eq.~(\ref{partition}) involves the phase $\phi (\tau ,\vec{r})$
and the charge integer fields $n(\tau ,\vec{r})$ as statistical
variables, in a three dimensional lattice with two spatial
dimensions $L_{x}\times L_{y}$,  and one imaginary time dimension,
$L_{\tau }$. The angular phases $\phi (\tau ,\vec{r })\in \lbrack
0,2\pi ]$ are defined at the nodes of the lattice with periodic
boundary conditions  in the  $x$ and $y$  space dimensions. The
integer fields, $n(\tau ,\vec{r})$,  lie in the bonds between two
consecutive nodes along the imaginary time axis $\tau ,$ and they
can take any integer values. The quantization condition in $\Xi $
is imposed in terms of
the periodic boundary conditions $\phi (L_{\tau },\vec{r}%
)=\phi (0,\vec{r})$. The  $L_{\tau }\rightarrow \infty $ limit has
been formally taken to replace the sum over imaginary time by
its integral. This $\Xi $  representation  is amenable to
numerical calculations in comparison to its operator
representation.

The arrays can be in a superconductor, insulator, or normal
states, depending on the values of $T$, $\alpha $ and $C_{{\rm
int}}$. To characterize the superconducting or normal state
behavior we calculated the helicity modulus, $\Upsilon $, for each
array,
\begin{equation}
\Upsilon =%
{\displaystyle{\frac{ \partial ^{2}{\cal F} }{ \partial {\rm k}^{2}}}}%
\Big|_{{\rm k=0}}.  \label{helicity}
\end{equation}
$\Upsilon $ measures the energy needed to carry out a phase twist
between the boundaries of the array along the $\vec{k}$ direction.
The helicity modulus is proportional to the superfluid density per
unit mass $\rho _{{\rm s}}$,
\begin{equation}
\rho _{{\rm s}}(T)=%
{\displaystyle{\frac{1} {V}}}%
\Big(%
{\displaystyle{\frac{ma }{ \hbar} }}%
\Big)\Upsilon (T).  \label{superfluid}
\end{equation}
Here $a$ is the lattice spacing, $m$ the Cooper pair mass, and $V$
the volume. Combining Eqs.~(\ref{partition})
and~(\ref{helicity}), we obtain the path integral
representation of the helicity modulus when the twist is along the
$\vec{x}$ axis~\cite{rojas96},
\begin{eqnarray}
{\displaystyle{\frac{1} {E_{J_{\nu }}L_{x}L_{y}}}}%
\Upsilon _{\nu }^{{\rm x}}(T)=%
{\displaystyle{1 \over L_{x}L_{y}L_{\tau }}}%
\Bigg[\Big<\sum_{\tau =0}^{L_{\tau }-1}\sum_{\vec{r_{\nu }}}\cos \Big[\phi
(\tau ,\vec{r_{\nu }})-\phi (\tau ,\vec{r_{\nu }}+\hat{x})\Big]\Big>\cr-%
{\dfrac{E_{J_{\nu }}\beta }{L_{\tau }}}%
{\displaystyle{E_{J_{\nu }}\beta  \over L_{\tau }}}%
\Bigg\lbrace\Big<\Big(\sum_{\tau =0}^{L_{\tau }-1}\sum_{\vec{r_{\nu }}}\sin %
\Big[\phi (\tau ,\vec{r_{\nu }})-\phi (\tau ,\vec{r_{\nu }}+\hat{x})\Big]%
\Big)^{2}\Big>\cr-\Big<\sum_{\tau =0}^{L_{\tau }-1}\sum_{\vec{r_{\nu }}}\sin %
\Big[\phi (\tau ,\vec{r_{\nu }})-\phi (\tau ,\vec{r_{\nu }}+\hat{x})\Big]%
\Big>^{2}\Bigg\rbrace\Bigg]\label{upsilonx}.
\end{eqnarray}
The charge coherence in the arrays is determined by the inverse
dielectric constant $\varepsilon ^{-1}$, defined as,
\begin{equation}
\epsilon ^{-1}={\rm lim}_{\vec{q}\rightarrow 0}\Big[1-%
{\displaystyle{Q^{2} \over k_{B}T}}%
{\displaystyle{1 \over {\bf C}(\vec{q})}}%
\big<n(\vec{q})n(-\vec{q})\big>\Big].  \label{dielectric}
\end{equation}
Combining Eqs.~(\ref{partition}) and~(\ref{dielectric}), as well as
Fourier transforming the capacitance matrix and the charge number
operator~\cite{rojas96}, we can write the path integral
representation of the correlation function $\big<n(\vec{r_{1}})n(\vec{r_{2}})%
\big>$ as,
\begin{equation}
\Big<n(\vec{r_{1}})n(\vec{r_{2}})\Big>=%
{\displaystyle{1 \over \beta Q^{2}}}%
{\bf C}(\vec{r_{1}},\vec{r_{2}})+\Big(%
{\displaystyle{2\pi  \over \beta L_{\tau }}}%
\Big)^{2}\sum_{\vec{r_{3}},\vec{r_{4}}}{\bf C}(\vec{r_{1}},\vec{r_{3}}){\bf C%
}(\vec{r_{2}},\vec{r_{4}})\Big\langle m(\vec{r_{3}}),m(\vec{r_{4}})%
\Big\rangle.
\end{equation}
Substituting this result in Eq.(\ref{dielectric}) yields the
inverse dielectric constant expression
\begin{equation}
\epsilon ^{-1}={\rm lim}_{\vec{k}\rightarrow 0}\Bigg[%
{\displaystyle{\big(2\pi \big)^{2} \over \beta Q^{2}}}%
{\bf C}(\vec{k})\big<\big|m(\vec{k})\big|^{2}\big>\Bigg].
\label{epsilon_path}
\end{equation}
In these equations the path integral representation of the integer fields $m(%
\vec{r})^{\prime }s$ is $m(\vec{r})=\sum_{\tau =0}^{L_{\tau }-1}m(\tau ,\vec{%
r})$. It is important to note that the path integral representation of $%
\epsilon ^{-1}$ in Eq.~(\ref{epsilon_path}) is not exactly the
inverse dielectric function of a gas of Cooper pairs, since it
depends on the discretization of the imaginary time axis in
$L_{\tau }$ slices. Nonetheless, we expect that it contains most
of the relevant information of the dielectric properties of the
gas of Cooper pairs in the arrays.

\section{Results and discussion}

\label{results-discussion}

\subsection{Parameter values in the simulations\label{simulations}}

To carry out the Monte Carlo moves in the phases we used the
standard Metropolis algorithm. To speed up the calculations, we
replace the $U(1)$ continuous symmetry in the phases by a discrete
${\bf Z}_{N}$ subgroup \cite{jacobs84}. Using $N=5000$ has proved
to yield good results. Discretizing the phases has the advantage
of using integer arithmetic that allows building cosine tables for
the Boltzmann factors in the partition function. This
simplification cannot be used for the integer fields, except when
$C_{{\rm m}}=0 $, in which case the  $m(\tau ,\vec{r})$ fields can
be summed up exactly. This approach allows to build up  look up
tables that introduce an adequate effective
potential~\cite{rojas96}.

Once the system reaches thermal equilibrium after $N_{{\rm ther}}$
(between $10^{3}$ and $10^{4}$) MC sweeps, the thermodynamic
averages of interest were calculated.  A measure was taken after
$N_{sweeps}$ passes through the arrays updating the phases, and
$M_{sweeps}$ passes updating the integer fields. In the
semi-classical limit, $\alpha \ll 1$, and we
needed to consider $N_{sweeps}=1$ and $M_{sweeps}=1$ at high temperatures ($%
T>0.4$),  and $N_{sweeps}=4$ and $M_{sweeps}=4$ at low temperatures ($T<0.4$).
In the quantum limit, $\alpha >1$, we took $N_{sweeps}=4$ and $M_{sweeps}=4$,
at high temperatures, and $N_{sweeps}=10$ and $M_{sweeps}=10$ at low
temperatures. These parameter values were chosen to minimize the
decorrelation times due to the long range charge interactions. Proceeding in
this way we carried out averages over 16384 MC steps at high temperatures and 32768
at low temperatures. Error bars were calculated using the biased reduction
method described in reference~\cite{jacknife}.

We  studied the helicity modulus and the inverse dielectric
constant for each array as a function of temperature, $\alpha $, and the
ratio of the inter-layer self capacitance to the intralayer mutual
capacitance, $%
{\displaystyle{C_{{\rm int}} \over C_{{\rm m}}}}%
$. The reduced temperature was varied in the range  $[0.05,1.0]$ in
0.05 steps at high $T^{\prime }s$ and 0.025 steps at low $T^{\prime }s$. To
study the transition between the semi-classical and quantum states, the
quantum parameter in one array was kept fixed at $\alpha _{1}=0.5$ while  we
varied the quantum parameter of the second array taking values, $\alpha
_{2}=0.5,1.0,1.25,1.50,2.0,2.5,3.0,$ and 4.0. In addition, the capacitance
ratios were integer multiples of the Delft's experimental parameters \cite
{herre92}: $%
{\displaystyle{C_{{\rm int}} \over C_{{\rm m}}}}%
=\kappa \times 0.26087$, with $\kappa =$ 1,2,3,4, and 5. The values chosen for
this ratio allowed us to study the effects of the capacitive coupling between the
arrays,  going from weak coupling,  to the strong
coupling regime when $\kappa \geq 3$. The simulations were carried out in
lattice sizes $L_{x}\times L_{y}=32\times 32$ and $L_{\tau }=32$, for
$0.5\leq \alpha \leq 2.0$,  and $L_{\tau }$ up to 96 for higher values
 $2.0<\alpha \leq 4.0$. The larger the value $\alpha$
the larger the imaginary time axis had to be. For the $L_{\tau }$
values chosen  we obtained reliable results, with negligible
finite size effects along the imaginary time axis.

\subsection{Results for $ \Upsilon $  and $ \epsilon^{-1} $}

\label{results} In this section we present and discuss the results
for $ \Upsilon_{1,2} $ and $ \varepsilon^{-1}_{1,2}$ for different
values of the quantum parameter $\alpha _{2}$ while  $\alpha
_{1}=0.5$ was  kept fixed. We start considering the case when both
arrays are in the semi-classical regime, $\alpha _{1}=0.5$ and
$\alpha _{2}=1.0$ and $0.26087\leq C_{{\rm int}} \leq 1.30435$.
The corresponding results are not shown explicitly here, however,
we briefly describe them to compare, where appropriate, with
previous studies. We found that at low temperature
$\Upsilon_{1,2}$ are finite --SC phase-- with  $\Upsilon_{1} >
\Upsilon_{2}$, with small thermal fluctuations.  Both quantities
decreased monotonically down to zero as temperature increased. For
$T\geq T_{\rm sc}(\alpha_{1,2}) $ they  were equal to zero
indicating  that the arrays were in the normal phase. The
superconductor to normal (SC-N) transition temperature,
$T_{SC}(\alpha)$, shifted downwards as the charging energy
increased. Around this transition temperature the fluctuations of
$ \Upsilon$ became  larger. Since the arrays were in the
semiclassical regime the charges did not play an important role in
the behavior of the system, except for the small decrease in phase
coherence,  and we obtained, $ \varepsilon^{-1}_{1,2}=0$, in the
whole temperature range. These results for the two coupled arrays
are in agreement with previous studies for one
array~\cite{jacobs84,jacobs87,jacobs88,rojas96}. The results
described here  indicate that $T_{SC}(\alpha)$ was weakly affected
by the interlayer capacitive coupling.

\begin{figure}[t]
\begin{center}
\includegraphics[width=5.5in,height=5.0in,keepaspectratio=true]
{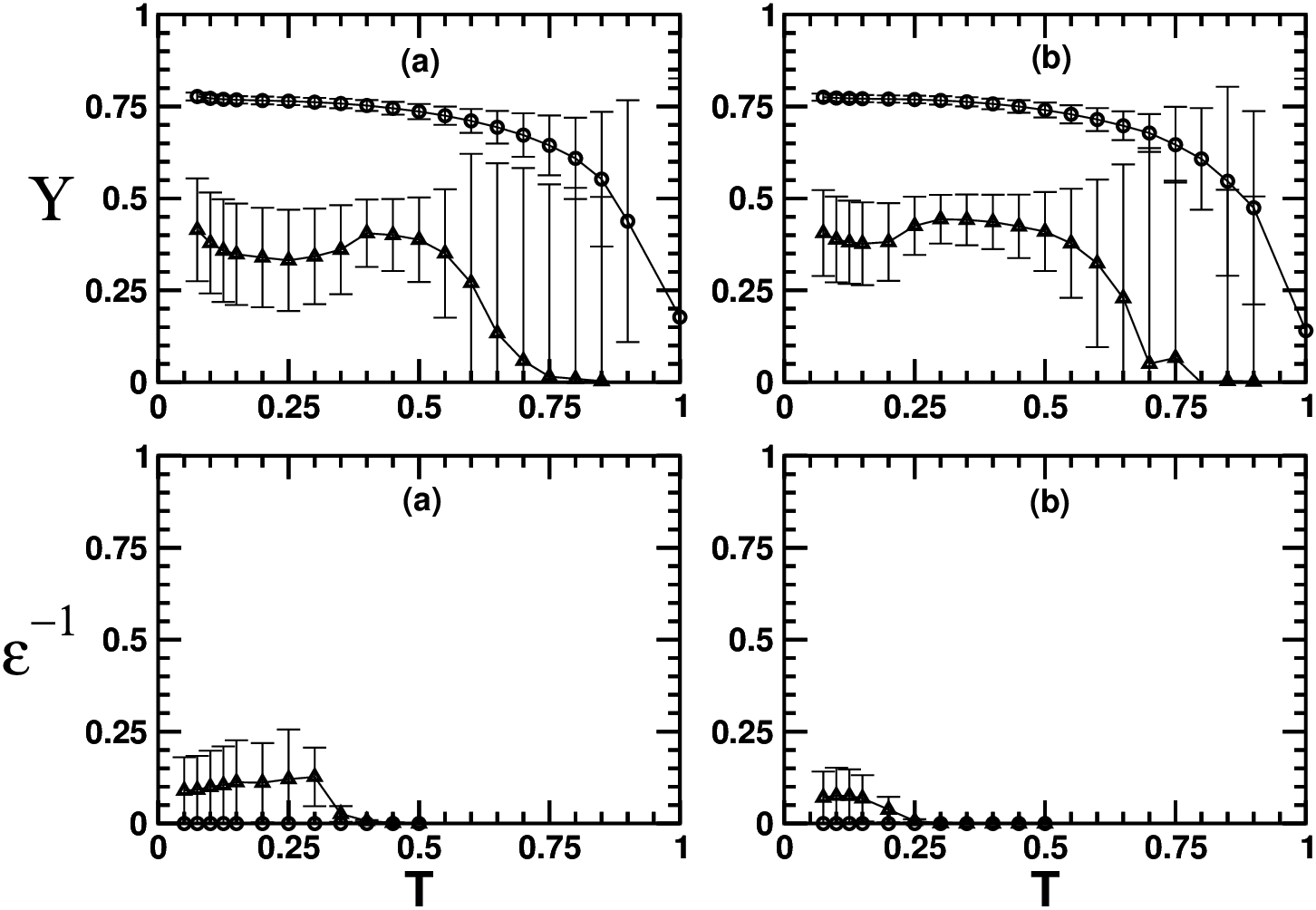}
\end{center}
\caption{$ \Upsilon $  and $ \epsilon^{-1}$ as a function of
temperature. $ \alpha_{1}=0.5 $ (circles), for the semiclassical
array, and $ \alpha_{2}=2.5 $ (triangles), for the quantum array.
The values of  $\frac{C_{\rm int}}{C_{\rm m}} $  are (a)
$0.52174$
 and (b) $0.78261$.}
\label{quantum1a}
\end{figure}

\begin{figure}[t]
\begin{center}
\includegraphics[width=5.5in,height=5.0in,keepaspectratio=true]
{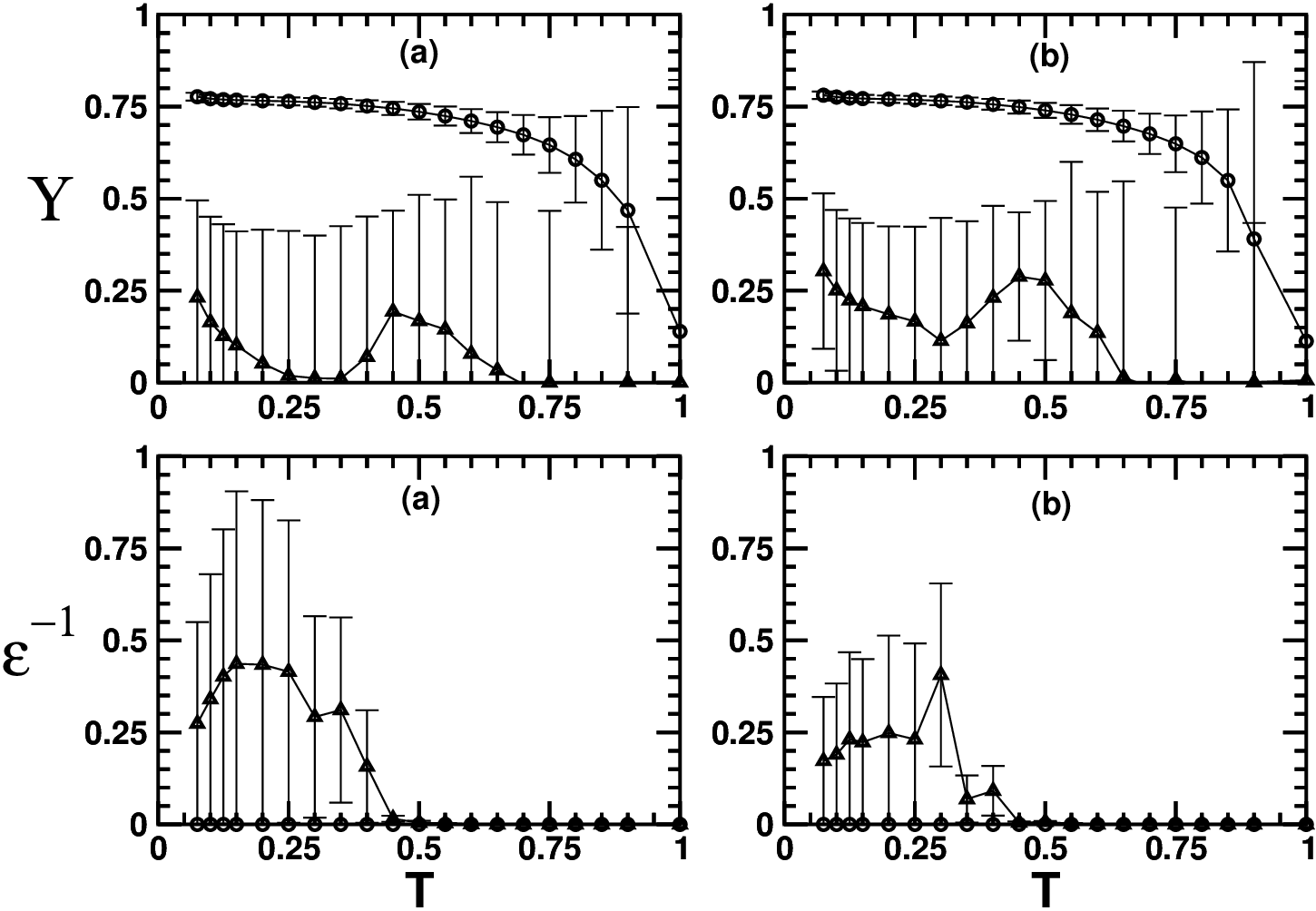}
\end{center}
\caption{Same as in Fig.~\ref{quantum1a} with  $ \alpha_{2}=3.0 $ and
$\frac{C_{\rm int}}{C_{\rm m}} =$  (a) 0.52174  and (b) 0.78261.}
\label{quantum2a}
\end{figure}

\begin{figure}[t]
\begin{center}
\includegraphics[width=5.5in,height=5.0in,keepaspectratio=true]
{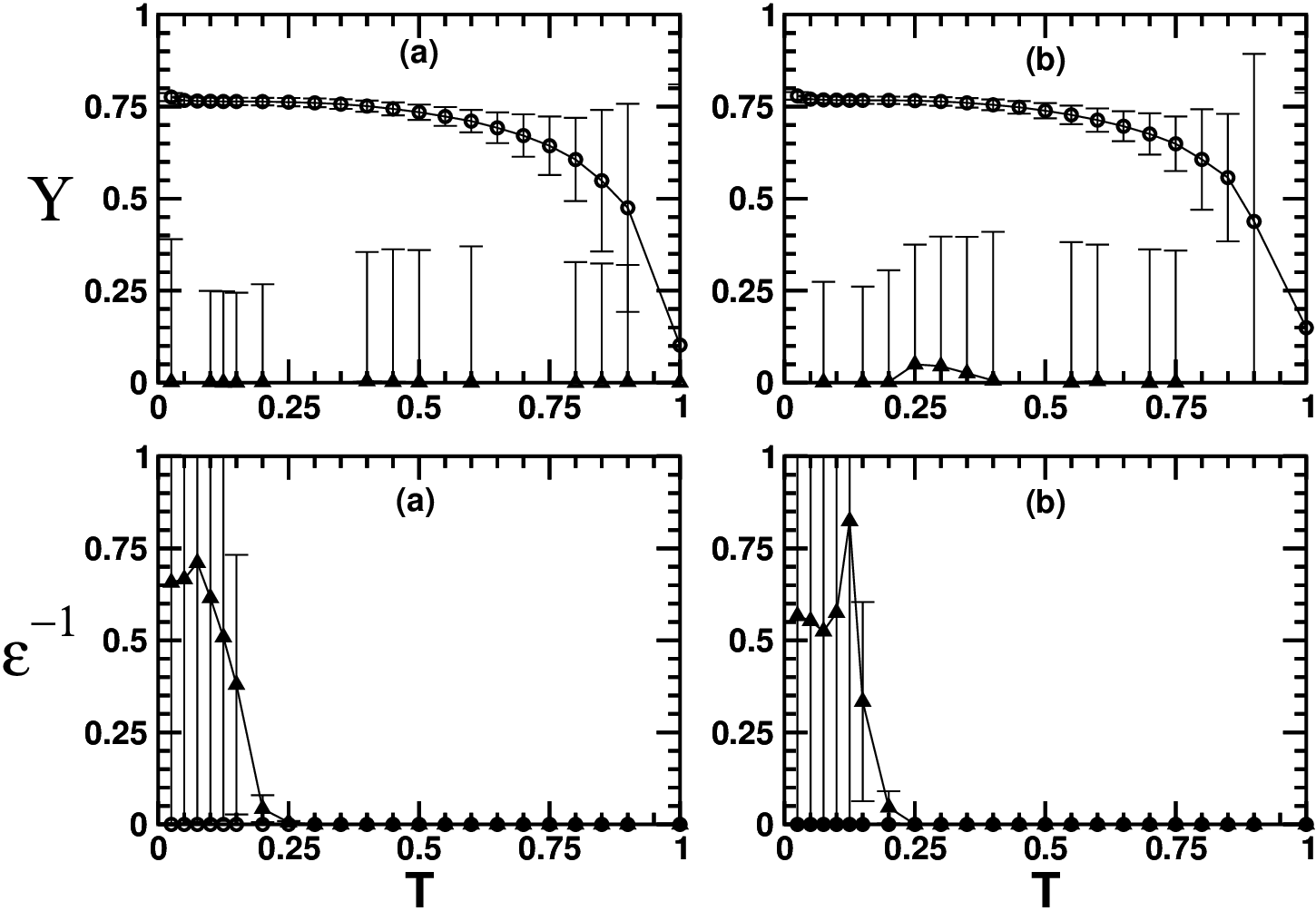}
\end{center}
\caption{Same as in Fig.~\ref{quantum1a} with
$ \alpha_{2}=4.0 $  and  $ \frac{C_{\rm int}}{C_{\rm m}} =$ (a) 0.52174
and (b) 0.78261.}
\label{quantum3a}
\end{figure}

\begin{figure}[b]
\begin{center}
\includegraphics[width=5.5in,height=5.0in,keepaspectratio=true]
{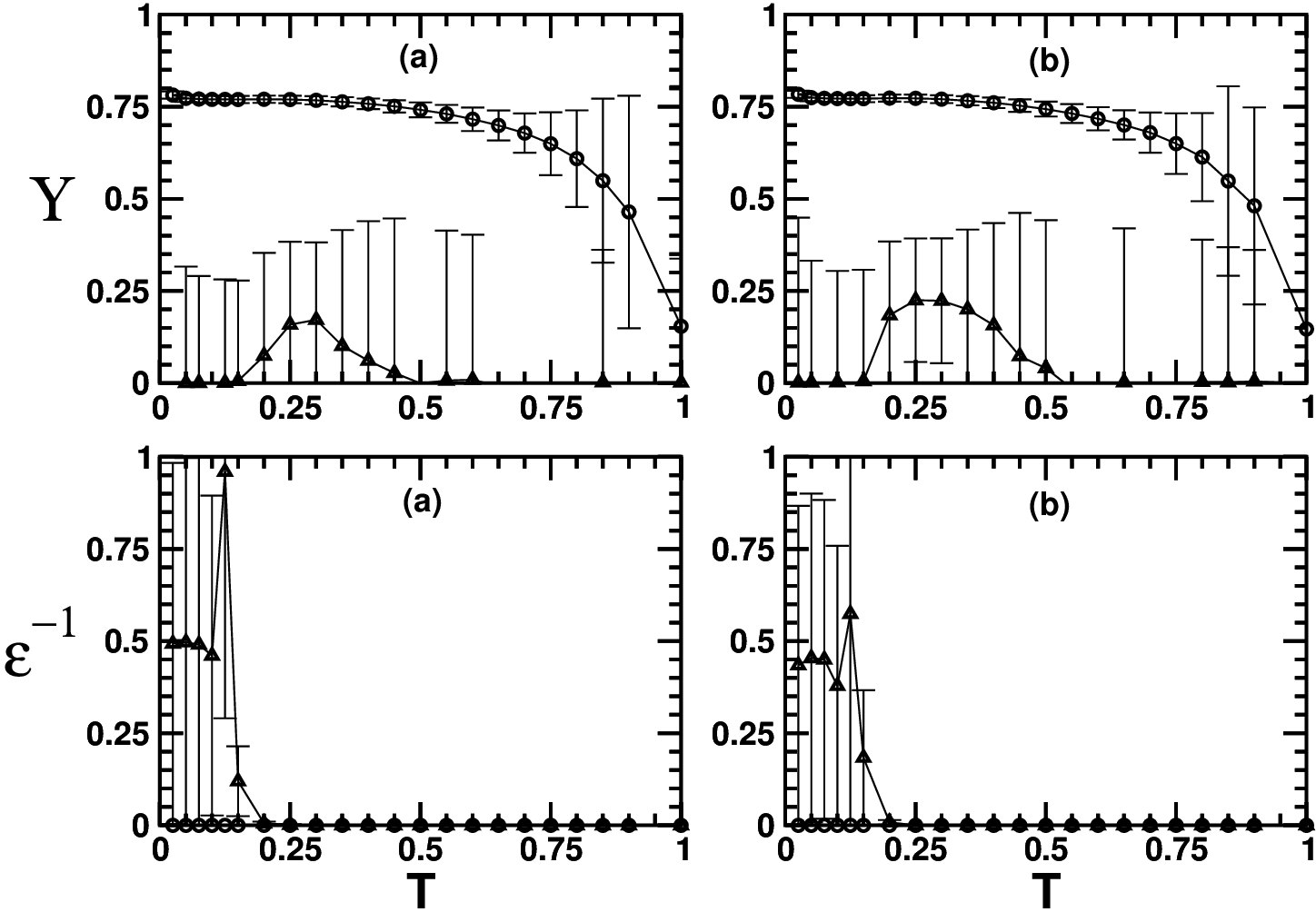}
\end{center}
\caption{Same as in Fig.~\ref{quantum3a} with
$ \frac{C_{\rm int}}{C_{\rm m}}$ (a) 1.04348  and (b) 1.30435.}
\label{quantum3b}
\end{figure}

In Fig.~\ref{quantum1a} we show the helicity modulus --upper panels-- and
the inverse dielectric constant --lower panels-- of each array as a function
of temperature when $ \alpha_{1}=0.5 $ and $\alpha _{2}=2.5$, for low and high values of
 $C_{{\rm int}} \over C_{{\rm m}}$.   The dependence of the SC-N phase boundary
$T_{SC}(\alpha)$   on $ C_{{\rm int}} $ will be shown and discussed later.
 In figure~\ref{quantum1a} we see that the
second array starts to develop  charge coherence: At very low
temperatures the inverse dielectric constant of 1, although small,
it has increased while the phase coherence of the second array
decreased,  as shown by the reduced value of the helicity modulus.
This behavior confirms the dual behavior between vortices in one
array and charges in the other one, due to their {\it gauge
interaction},  as predicted in reference~ \cite{jose98}. 
There, an effective action was written in terms of four interacting imaginary     
time Coulomb-like gases. The effective Hamiltonian was shown to be dually         
symmetric between charge and vorticity with complicated kernels. In the simplest   
case, where one array has one vortex and the other one charge,  their interaction  
has a minimal gauge coupling that is proportional to the site coupling capacitance 
between the arrays.                                                                                        
As a consequence of this {\it gauge capacitive interaction,} the
fluctuations in the helicity modulus and the inverse dielectric
constant are much stronger than those observed in the
semi-classical limit in a single array. Nonetheless, the quantum
fluctuations in the second array are not sufficiently strong to
significantly affect the semi-classical behavior of the first
array. This indicates that the {\it gauge interaction} between
vortex and charge Coulomb gases is weak.
On the other hand, it also found that as $ C_{\rm int} $ increases
the charge coherence of the second array decreases, restoring its
phase coherence. In addition there appears to be a reentrant phase
transition in the second array at low temperatures, as suggested
by the concave curvature of its helicity modulus and the slight
convex curvature of its inverse dielectric constant. This picture
becomes more evident when increasing further the charging energy
in the second array, to  $\alpha _{2}=3.0$, while  keeping $\alpha
_{1}=0.5$. The results are  shown in Fig.~\ref{quantum2a}. When
$\frac{C_{{\rm int}}}{C_{{\rm m}}}=0.52174$, as the system cools
down, array 2 undergoes a SC-N  phase transition at $T\simeq
0.35$, followed by a reentrant N-SC transition at $T\simeq 0.25 $.
In addition, we also observe that at $T\simeq 0.45$ the charge
coherence increases significantly and then decreases at $T\simeq
0.15$. This suggests an insulating to normal reentrant transition.
Once again, this is a manifestation of the dual character between
the charge and phase degrees of freedom in the two arrays as well
as the {\it gauge interaction} between the vortex and charge
Coulomb gases. As the interaction capacitance increases further to
$\frac{C_{{\rm int}}}{C_{{\rm m}}}=0.78261$, the reentrant
transition in the phase degrees of freedom is less evident, but it
is still present. A similar situation  is observed in the charges
degrees of freedom. Increasing further the arrays interaction
capacitance improves the phase coherence of the arrays at the
expense of their charge coherence. Note that the thermal
fluctuations are strong in the phase and in the charge degrees of
freedom for the array parameters considered here. In spite of the
capacitive interaction between arrays 1 and 2, the behavior in
array 2 does not affect significantly  the behavior of array 1.
When the quantum parameter of array 2 is sufficiently large,
$\alpha _{2}=4.0$, the quantum effects become dominant, and the
charge degrees of freedom destroy  the phase coherence when
$\frac{C_{{\rm int}}}{C_{{\rm m}}} =0.52174$ and 0.78261. This
behavior is shown in figure~\ref{quantum3a}. For this value of
$\alpha _{2}$ we needed to consider $L_{\tau }=96$ for array 2 to
get meaningful results. We found no transition at all in the
phases, but we do get a sharp N-I  transition for the charges,
although the fluctuations in those degrees of freedom are rather
large. Increasing  $ C_{\rm int} $ further leads to a different
scenario. For $\frac{C_{{\rm int}}}{C_{{\rm m}}} =1.04348 $, a SC
phase appears in the temperature range $0.25\leq T\leq 0.45$,
followed by an insulating phase at lower temperatures. For
$\frac{C_{{\rm int}}}{C_{{\rm m}}}=1.30435$ we obtain a similar
situation. In this case the phase coherence is greater than for
the previous values of the interaction capacitance parameter.
These results are shown  in figure~\ref{quantum3b}.

\begin{figure}[t]
\begin{center}
\includegraphics[width=5.5in,height=5.0in,keepaspectratio=true]
{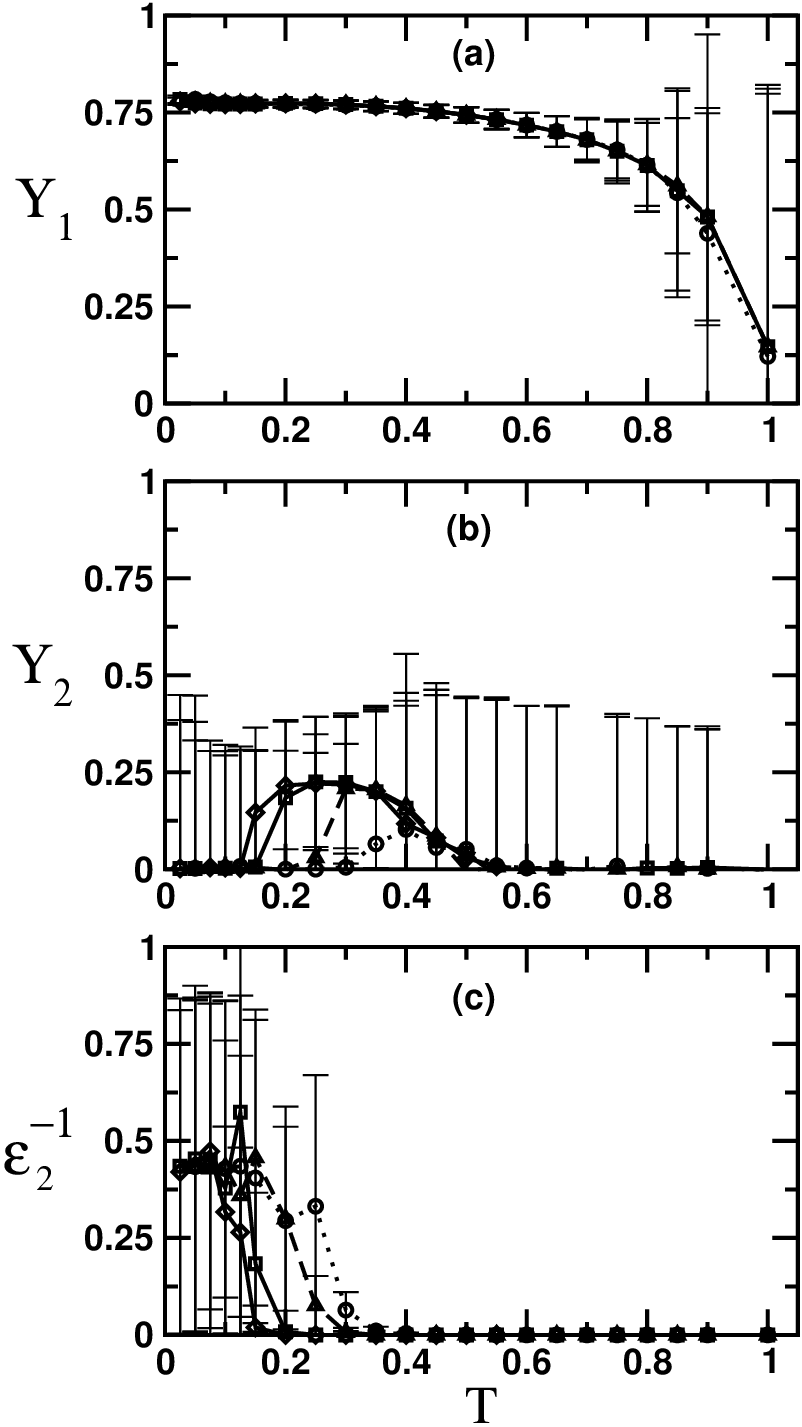}
\end{center}
\caption{$ L_{\tau} $ finite size behavior of $ \Upsilon_{1,2}$
and $ \epsilon^{-1}_{2} $  for $ L_{\tau} =$48 (circles), $
L_{\tau} =$64 (triangles),  $ L_{\tau} =$96 (squares), and 
$ L_{\tau} =$128 (diamonds). 
The interaction between arrays is $ \frac{C_{\rm int}}{C_{\rm m}}
=1.30435$. (a) semiclassical array ($ \alpha_{1}=0.5 $), (b) and
(c) quantum array ($ \alpha_{2}=4.0 $). $ \epsilon^{-1}_{1} $ is
equal to zero in the whole temperature range and is not shown.}
\label{finite-lt}
\end{figure}

As  pointed out in section~\ref{model}, when  the quantum
parameter, $ \alpha \geq 1$, the numerical evaluation of the path
integral representation of $ \varepsilon^{-1} $
--Eq.~(\ref{epsilon_path})--  depends importantly on the time
slice discretization of the imaginary time axis, that is, there
are  finite size effects.  To ascertain this situation  we
calculated  $\Upsilon (T)$ and $ \epsilon^{-1} (T)$   for four
different imaginary time axis lengths, $ L_{\tau}=48,64,96 $, and 128.
In figures~\ref{finite-lt}(a),(b) we show the results of these
calculations for $\Upsilon_{i} (T)$, $ i=1,2 $  when
$\frac{C_{{\rm int}}}{C_{{\rm m}}}=1.30435$ with $ \alpha_{1}
=0.5$, and $ \alpha_{2} =4$. We see that $ \Upsilon_{1} (T) $
shows no finite size effects  since the data for
different values of $ L_{\tau}$ for each $T$ fall on top of each
other. There are very small finite size effects for those
temperatures close to the SC-N transition temperature. However, 
$\Upsilon_{2} (T) $ shows important finite size effects in the
temperature region where the reentrant transition sets in.  This
reentrance becomes better defined, and its behavior becomes smoother
for $ L_{\tau}\geq 96$. Similarly, the  SC-I temperature shifts slightly to 
lower temperatures for $ L_{\tau}\geq 96$, while the N-SC
temperature remains about the same. On the other hand,  
$\epsilon^{-1}_{1} (T)=0$ in the whole temperature range
considered.  This is not unexpected since  array 1 is in the
semi-classical regime  ($ \alpha_{1} =0.5$), and the role of the
charge degrees of freedom is negligible. Because of this we do not
show $ \epsilon_{1}^{-1} $ in figures~\ref{finite-lt}.
Nonetheless, $ \epsilon^{-1}_{2} (T)$ is sensitive to the 
$L_{\tau}$  length  axis, as can be seen  in
figure~\ref{finite-lt}(c). In this case, we see that the
insulating phase sets in at the temperature boundary $ T_{SC-I}
\big(L_{\tau} \big)$. For $ L_{\tau}\geq 96$ it shifts slightly to 
lower temperatures. We also see  strong fluctuations in the inverse
dielectric constant at about  $ T_{SC-I}$. For  other values of
$\frac{C_{{\rm int}}}{C_{{\rm m}}}$, and $ \alpha_{2}$, the
results are less significant.
Although we did not see a clear saturation of 
$ \epsilon^{-1}_{2} (T)$ at the reentrance transition for the two 
largest values of $ L_{\tau}$ we were able to minimize the finite 
size effects along the imaginary time length. 
Carrying out calculations using larger imaginary time axis becomes 
very demanding from the computational point of view.

\subsection{Phase diagrams}

\label{phase-diagrams} Here we present the cumulative results from
the estimation of the critical temperatures from our
$\Upsilon(T,\alpha)$ and $\epsilon^{-1}(T,\alpha)$ calculations
that provide the phase boundaries for the SC-N and I-N phase
transitions. In array 1 the quantum parameter  $\alpha _{1}=0.5$,
is kept fixed, while $\alpha_2$ was varied in the interval
$1.0\leq \alpha _{2}\leq 4.0$, in $\Delta \alpha_2=0.5$ steps, for
each one of the values of $C_{\rm int}/C_{m}$. The transition
temperatures were estimated from the behavior of $\Upsilon$ and $
\varepsilon^{-1} $ as functions of temperature, for each value of
$ \alpha_{2} $. In the classical arrays, according to the
Berezinskii-Kosterlitz-Thouless (BKT) theory, the critical SC-N
temperature occurs at the intercept of the helicity modulus with a
straight line with slope $\frac{2}{\pi }$. In the quantum arrays
it has been shown experimentally  that in the limit
$C_{s}/C_{m}\rightarrow 0$ there is a crossover from a conducting
to an insulating phase~\cite{herre91,tighe93,delsing94}. Due to
the finite size of the experimental arrays there is a rounding of
the transitions, because the screening length is shorter than the
size of the arrays~\cite{herre92}. Theoretically it has been
argued that for any finite screening length the phase transition
is washed out even in arrays of infinite size~\cite{minnhagen93}.
This happens because in the BKT scenario the SC-N transition
crucially depends on the unscreened nature of the logarithmic
interaction between vortex pairs~\cite{bkt,jose77}. Nonetheless,
it has been recently shown~\cite{luca2003,luca2004} that even in
the regime of strong quantum fluctuations the data for $ \Upsilon
(T) $, at low temperatures, can be very well fitted  to the
Kosterlitz's renormalization group equations in a finite size
analysis of $ \Upsilon $. In spite of the theoretical and the
experimental differences in estimating the transition
temperatures, here we will use the BKT theory since we have not
carried out a detailed finite size analysis. When  there is a low
temperature reentrant phase transition, the transition temperature
is determined from the temperature at which the physical quantity
of interest vanishes, as is done with the arrays's electric
resistance in experiments. The error bars in the transition
temperatures were estimated taking into account the size of the
temperature variation step, $\Delta T$, in the calculations. Since
$\Delta T=0.05$, then the error bars in the transition
temperatures of the phase boundaries are $\delta T_{{\rm
c}}=0.05$.

\begin{figure}[b]
\begin{center}
\includegraphics[width=5.5in,height=5.0in,keepaspectratio=true]
{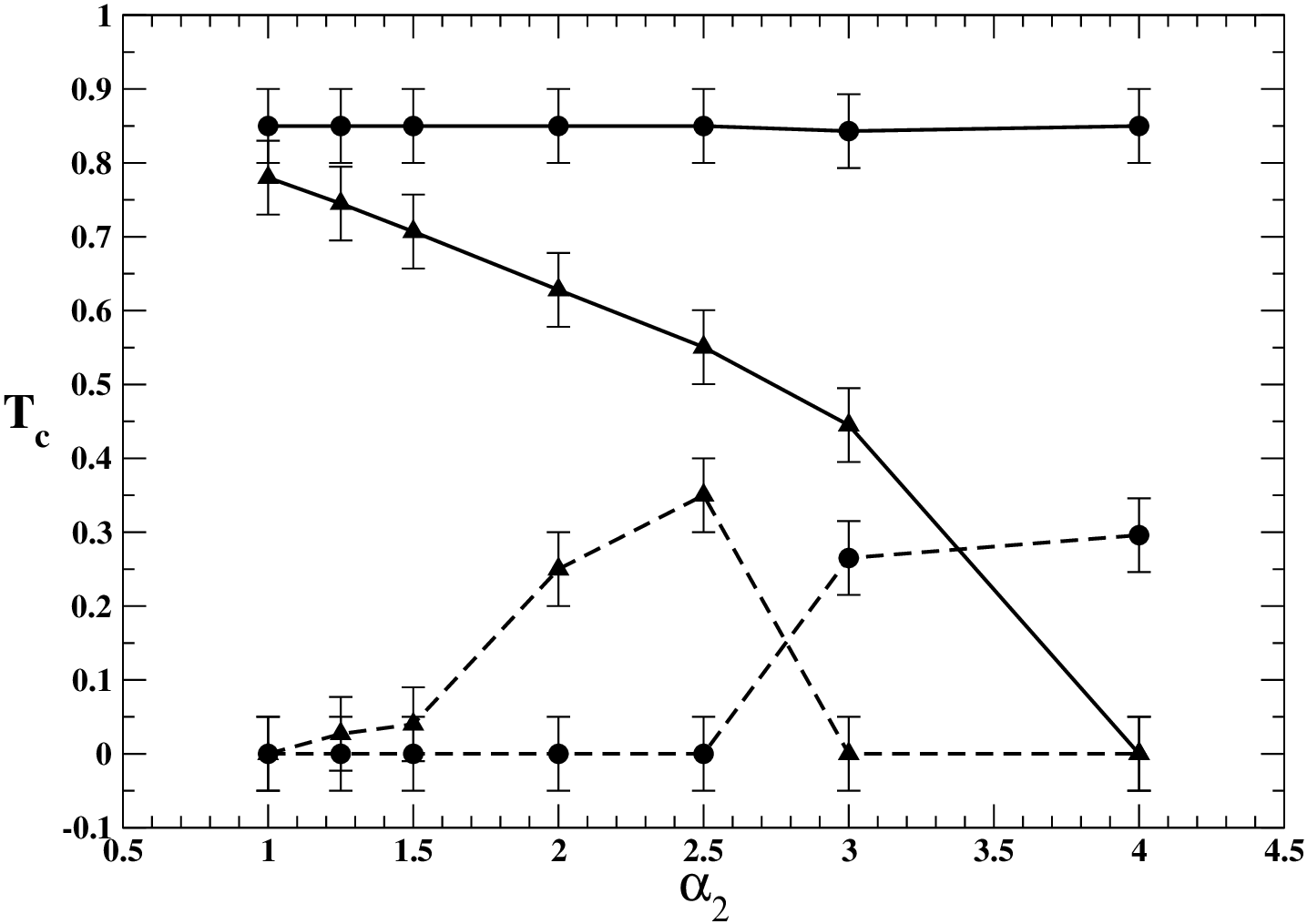}
\end{center}
\caption{Estimated transition temperatures of each array
versus $ \alpha_{2} $ for  $ \frac{C_{\rm int}}{C_{\rm m}} =0.52174$.
Array 1 ({\Large $\bullet$}), array 2 ($\blacktriangle$).  The solid lines
represent the SC-N phase boundary while the dashed lines the
I-N phase boundary. }
\label{phase_1}
\end{figure}

\begin{figure}[b]
\begin{center}
\includegraphics[width=5.5in,height=5.0in,keepaspectratio=true]
{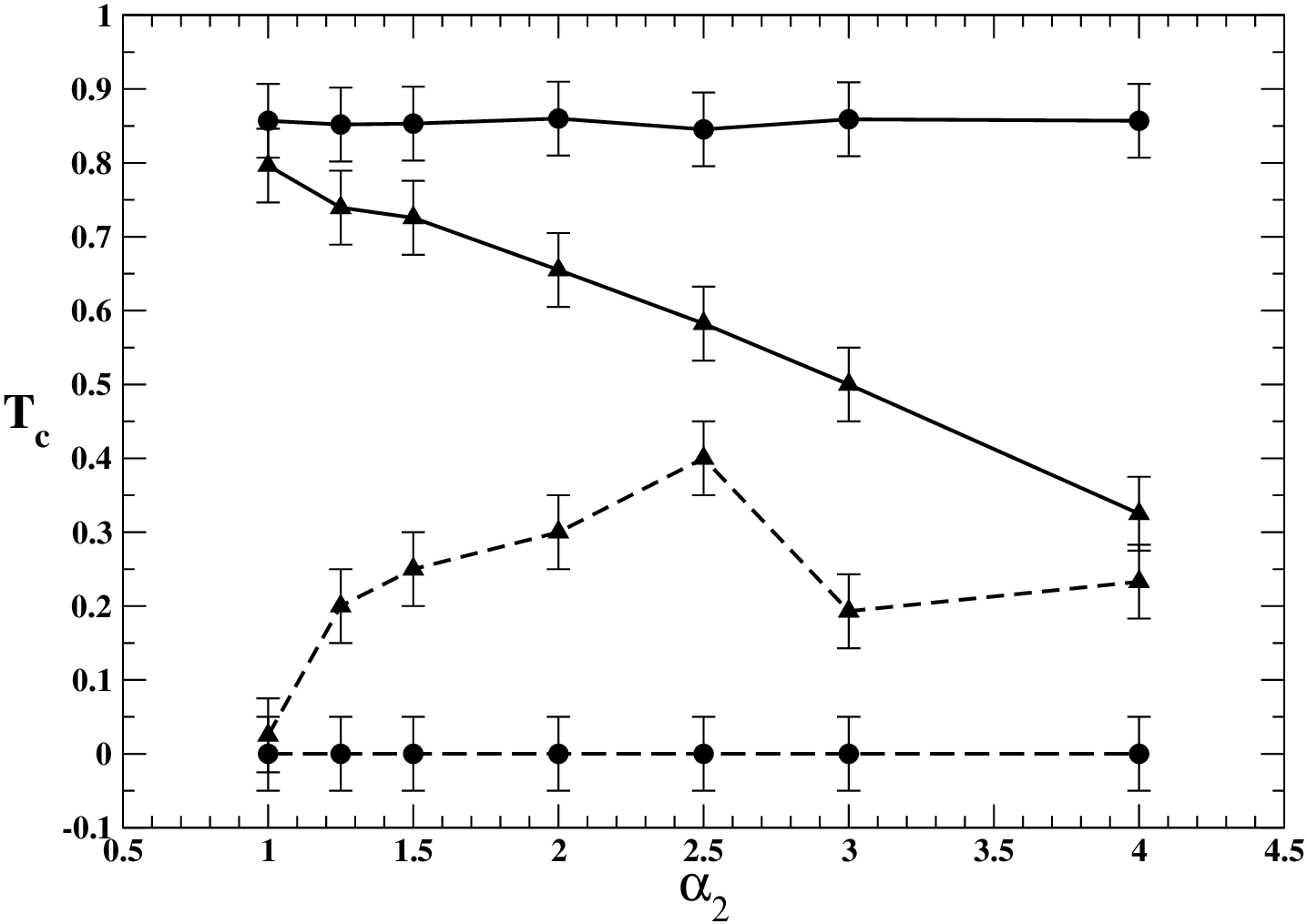}
\end{center}
\caption{Same as in Fig.~\ref{phase_1} with
$ \frac{C_{\rm int}}{C_{\rm m}} =1.04348$.}
\label{phase_3}
\end{figure}

In figure \ref{phase_1} we show the phase diagrams $T_{c}$ versus
$\alpha _{2}$ for the phase and charge degrees of freedom for
arrays 1 and 2, when $C_{\rm int}/C_{m}=$0.52174. We see that the
SC-N temperature boundary of the semiclassical array 1 remains
almost unchanged as a function of $\alpha _{2}$, with $T_{{\rm
c}}\simeq 0.85$. The SC-N phase boundary of array 2, however,
changes significantly as $\alpha _{2}$ increases. The transition
temperature decreases monotonically and it extrapolates  to zero
at $\alpha_{2}=4$. At this $ \alpha_{2} $ value the second array
is fully in the quantum regime. Similar results were obtained  for
$C_{\rm int}/C_{m}=$0.26087. It was also found that  the I-N phase
boundaries of the arrays  are qualitatively similar for
 $1\leq \alpha _{2}\leq 2.5$ with $C_{\rm int}/C_{m}=$ 0.26087 and 0.52174.
That is,  for array 1 $ T_{\rm IC}(\alpha_{2})=0 $,  and  for
array 2 the I-N transition temperature increases monotonically and
reaches a maximum at $\alpha _{2}=2.5$. However,  when array 2 is
in the quantum regime, $ \alpha_{2} > 2.5$, the situation is
different. There we found that  the I-N boundary   of array 1 is
equal to zero when $C_{\rm int}/C_{m}=$ 0.26087. Increasing this
quantity up to  0.52174, yields  a nonzero and monotonically
increasing $ T_{\rm IC}(\alpha_{2}) $. For array 2 with $C_{\rm
int}/C_{m}=$0.26087 we found that the $ T_{\rm IC}(\alpha_{2})$
decreases slowly and for $ 3 \leq \alpha_{2} \leq 4 $ it becomes
constant, $ T_{\rm IC} \simeq 0.3 $.  For larger values of the
interaction parameter, $C_{\rm int}/C_{m}=$ 0.52174,  the I-N
phase boundary decreases down to zero for $3.0\leq \alpha _{2}\leq
4.0$, indicating the existence of a reentrant {\it novel} phase
transition. The results for $C_{\rm int}/C_{m}=$0.52174  are
plotted in figure~\ref{phase_1}.
For $C_{\rm int}/C_{\rm m}=0.78261$,  the SC-N phase diagram for
array 2, and the I-N phase diagram of both arrays, are modified
slightly.  The $ T_{\rm SC-N} $ boundary of array  2 decreases at
about the same rate when $1.0\leq \alpha _{2}\leq 3.0$ for the
values of $C_{\rm int}/C_{\rm m}$ given above. Nonetheless, the
transition temperature for $\alpha _{2}=4.0$ is not zero, instead,
it moves upwards. $ T_{\rm SC-N} $ may be approximated by a
straight line for the whole interval of $\alpha _{2}$ considered
here. The I-N boundary, it increases monotonically from $\alpha
_{2}=1.25$ up to $\alpha _{2}=2.0$, where it reaches a maximum and
then slowly decreases appearing to reach a minimum at $\alpha
_{2}=3.0$. Above this value it increases slowly. These results are
not shown here since they are similar to those in
Fig.~\ref{phase_3}.
A further increase in the coupling capacitance between the arrays
to, $C_{\rm int}/C_{\rm m}=1.04348$, does not change the SC-N and
the I-N boundaries of  array 1, but it changes the SC-N phase
diagram for array 2, and the I-N phase diagram of both arrays. The
SC-N transition temperature of array 2 decreases linearly as a
function of $\alpha _{2}$, while the I-N boundary increases
monotonically in the temperature range $1.0\leq T\leq 2.5$. It
reaches a maximum at $T=2.5$ and then decreases slowly up to
$\alpha _{2}=3.0$, where it has a minimum and then increases again
up to $\alpha _{2}=4.0$. Note that the position of the maximum is
shifted to higher values of $\alpha _{2}$, and the transition
temperature at $\alpha _{2}=4.0$ shifts downwards compared to the
previous value of $C_{\rm int}/C_{\rm m}$.
When $C_{\rm int}/C_{\rm m}$ is at the highest value considered
here, 1.30305, the SC-N boundaries of arrays 1-2 become straight
lines. The former has zero slope and the latter shows a negative
slope, i.e. the SC-N transition temperature of array 2 decreases
linearly as $\alpha _{2}$ increases. On the other hand, the I-N
boundary of array 1 starts as a horizontal straight line for
$1\leq \alpha _{2}\leq 1.5$ then increases reaching a maximum at
$\alpha _{2}=2.5$, then decreasing monotonically down to zero at
$\alpha _{2}=4.0$. These results suggest, again,  the existence of
a {\it reentrant transitions} in array 2 when it is in the quantum
regime. In contrast, the I-N boundary of array 2 is a horizontal
line at $T=0$, for $1\leq \alpha _{2}\leq 3.0$, and $ T_{\rm I-N}
(\alpha)$ becomes  nonzero for $\alpha _{2}>3.0$.

The structure of the phase diagrams indicate that by increasing
the capacitive coupling between arrays and increasing the
$\alpha_2$ parameter one can induce novel reentrant phase
transitions not only in the phase degrees of freedom but also in
the charge freedoms. This is a situation that does not occur in
single JJA.


\section{Conclusions}

\label{conclusions}

We have carried out extensive path integral
quantum MC simulations of two capacitively coupled JJA with ultra
small junctions. One in the semiclassical limit and the other in
the quantum regime. We studied the behavior of the helicity
modulus and the inverse dielectric constant as a function of
temperature for different values of the interlayer capacitances.
When both arrays are in the semiclassical regime, regardless of
the interlayer coupling considered here, each array shows a SC-N
transition at finite temperatures. For these semiclassical arrays
the charge degrees of freedom contribution is negligible and there
is no I-N
transition at finite temperatures. As array 2 enters the quantum regime ($%
2.0\leq \alpha _{2}\leq 2.5$), a SC-N reentrant phase transition
appears and the fluctuations in $ \Upsilon_{2} $  and $
\epsilon^{-1} $ become significantly larger due to quantum
fluctuations as explained in references~\cite{luca2003,luca2004}.
At the same time the quantum array starts to develop charge
coherence with an finite temperature I-N transition. This scenario
occurred for all the values of the interlayer coupling considered
here. As array 2 becomes more quantum ($3.0\leq \alpha _{2}<4.0$),
and for all the interlayer couplings considered, {\it reentrant
transitions} occur not only in the phase degrees of freedom, but
also in the charges degrees of freedom. When $\alpha _{2}=4.0$,
the SC-N transition is washed out for values of the interlayer
coupling $C_{\rm int}/C_{\rm m}\leq 0.78261$, having only an I-N
transition. Increasing further the interlayer coupling yields a
SC-N reentrant phase transition at low temperatures for
$0.20\lesssim T\lesssim0.5$, together with the usual I-N phase
transition for the charges. These scenarios can be understood as a
\ manifestation of the {\it gauge interactions between the phase
degrees of freedom and the charge degrees of freedom that are
present in each array}. The latter results depend on the values of
the quantum parameters. 
Finally, we showed a series of interesting                  
phase diagrams for both arrays where we found                
some resemblance to the reentrant-like behavior   observed  
experimentally by the Delft's group in 2D JJA with charging 
energy~\cite{herre92,herre96}.                              
\section{Acknowledgements}

GRS would like to acknowledge partial financial support from CONACYT-MEXICO
through contract 25298-E and from DAGAPA-UNAM contract IN110103.
JVJ thanks NSF for partial financial support.
\newpage


\newpage

\newpage

\newpage

\newpage

\newpage

\begin{figure}[t]
\begin{center}
\includegraphics[width=5.5in,height=5.0in,keepaspectratio=true]
{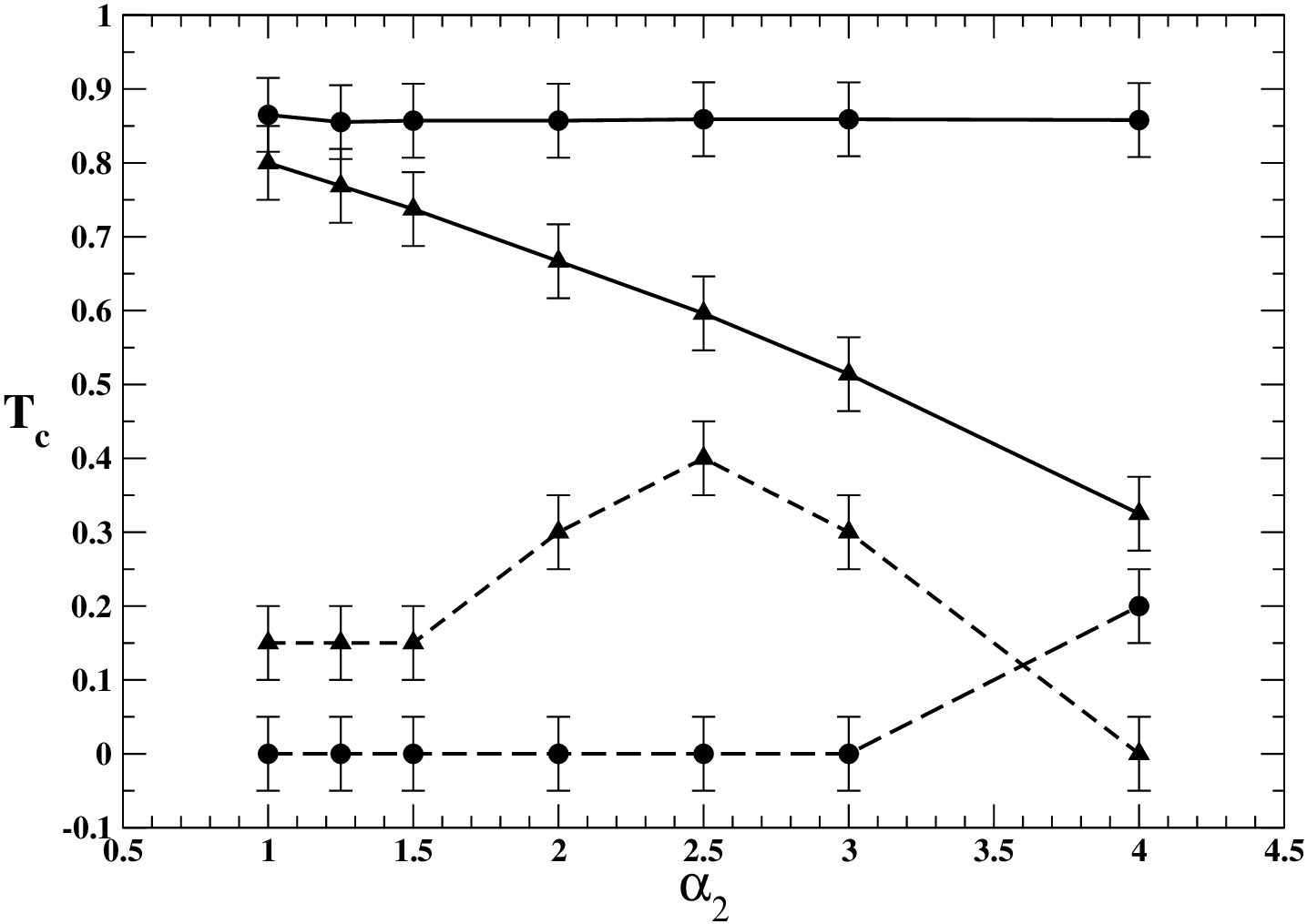}
\end{center}
\caption{Same as in Fig.~\ref{phase_1} with
$ \frac{C_{\rm int}}{C_{\rm m}} =1.30345$.}
\label{phase_4}
\end{figure}

\end{document}